\documentclass[aps,showpacs,preprint]{revtex4}
\usepackage{graphicx}
\begin{document}
\title{Iterative quantum state transfer along a chain of nuclear spin qubits
\footnote{Corresponding authors:
Jingfu Zhang, zhangjfu2000@yahoo.com, Jingfu@e3.physik.uni-dortmund.de;\\
Dieter Suter, Dieter.Suter@uni-dortmund.de }}
\author{
Jingfu Zhang, Nageswaran Rajendran, Xinhua Peng, and Dieter Suter }
\address{Fachbereich Physik, Universit$\ddot{a}$t Dortmund, 44221 Dortmund, Germany\\
}
\date{\today}

\begin{abstract}

Transferring quantum information between two qubits is a basic
requirement for many applications in quantum communication and
quantum information processing. In the iterative quantum state
transfer (IQST) proposed by D. Burgarth et al. [Phys. Rev. A 75,
062327 (2007)], this is achieved by a static spin chain and a
sequence of gate operations applied only to the receiving end of the
chain. The only requirement on the spin chain is that it transfers a
finite part of the input amplitude to the end of the chain, where
the gate operations accumulate the information. For an appropriate
sequence of evolutions and gate operations, the fidelity of the
transfer can asymptotically approach unity. We demonstrate the
principle of operation of this transfer scheme by implementing it in
a nuclear magnetic resonance quantum information processor.
\end{abstract}

\maketitle
\section{Introduction}
Quantum state transfer (QST), i.e., the transfer of an arbitrary
quantum state $\alpha|0\rangle+\beta|1\rangle$ from one qubit to
another, is an important element in quantum computation and quantum
communication \cite{books,Bose03,PST,Bose,thesis}. The most direct
method to implement QST is based on SWAP operations \cite{swap}.
This approach consist of a series of SWAP operations between
neighboring qubits until the quantum state arrives at the target
qubit. In a general-purpose quantum register, these quantum gates
require the application of single- as well as two qubit operations.
For longer distances, the number of such operations can become quite
large; it may then be advantageous to rely on quantum teleportation
instead \cite{telep}, which requires fewer gate operations, but
shared entanglement between sender and receiver.

For specific systems, it is possible to transfer quantum information
without applying gate operations, but instead relying on a static
coupling network \cite{Bose03,PST}. The main difficulty with this
approach is the required precision with which the couplings have to
be realized in order to generate a transfer with high fidelity.

This requirement can be relaxed significantly, without compromising
the fidelity of the transfer, by applying gate operations to the
receiving end of the spin chain that effects the transfer
\cite{Bose}. The capability for applying such gate operations is not
an additional requirement, since such operations are required anyway
if the spin chain is to be used for communication between quantum
registers. This gate accumulates any amplitude of the initial state
that is transferred along the chain. The protocol allows one, in
principle, to obtain unit fidelity for the transfer, even if the
couplings along the chain have arbitrary fluctuations, as long as a
finite amplitude reaches the end of the chain. Obtaining a large
transfer amplitude requires multiple iterations, each of which
includes the evolution of the spin chain and the two-qubit gate
operation. The fidelity for transfer increases with the number of
the iterations and can approach $1$ asymptotically. Hence we refer
to this protocol as the iterative quantum state transfer (IQST). In
this paper we implement the protocol in an NMR quantum information
processor and demonstrate its basic feasibility.

\section{Iterative transfer algorithm}

\subsection{System}
\begin{figure}
\includegraphics[width=3in]{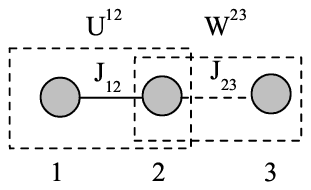}    
\caption{The spin chain including the target spin ($3$) used for
implementing the IQST. The XY- interactions in the spin chain,
denoted by the solid line, is always active,  while the XY-
interaction between spins $2$ and $3$, denoted by the dashed line,
can be switched on and off. $W^{23}$ denotes the end gate applied to
spins 2 and 3. $U^{12}$ denotes the evolution of spin chain.}
\label{cha}
\end{figure}

 We illustrate the
IQST proposed in Ref. \cite{Bose} using a system of three spins
coupled by Heisenberg XY- interactions, as shown in Figure \ref{cha}.
The spin chain
consists of spins $1$ and $2$, which are coupled by a constant
(time-independent) interaction.
Spin 3 is the target spin used to
receive the transferred quantum state. The interaction between spins
$2$ and $3$ can be switched on and off. Our purpose is to transfer
an arbitrary quantum state $\alpha|0\rangle+\beta|1\rangle$ from
spin $1$ to $3$, where $\alpha$ and $\beta$ are two complex numbers
normalized to $|\alpha|^2+|\beta|^2=1$.

The Hamiltonian of the the spin chain without the end qubit is
\begin{equation}\label{Ham12}
    H_{12}=\frac{1}{2}\pi
    J_{12}(\sigma_{x}^{1}\sigma_{x}^{2}+\sigma_{y}^{1}\sigma_{y}^{2}) ,
\end{equation}
where $J_{12}$ denotes the coupling strength.
The Hamiltonian of spins $2$ and $3$ is
\begin{equation}\label{Ham23}
    H_{23}(t)=\frac{1}{2}\pi J_{23}(t)(\sigma_{x}^{2}\sigma_{x}^{3}+\sigma_{y}^{2}\sigma_{y}^{3}) ,
\end{equation}
where $J_{23}(t)$ is $J_{23}$ when the interaction is
switched on and $0$ otherwise.

\subsection{IQST algorithm} \label{IQSTtheory}

The purpose of the IQST algorithm is the transfer of an arbitrary state
$\alpha|0\rangle+\beta|1\rangle$ from the start of the chain (qubit 1)
to the end (qubit 3).
We start the discussion by choosing as the initial state of the complete
3-qubit system the state $\alpha|000\rangle+\beta|100\rangle$,
i.e. a product state with spin $1$ in state
$\alpha|0\rangle+\beta|1\rangle$, and spins $2$ and $3$ in
$|0\rangle$.
Transferring the $\alpha|0\rangle$ part of the input
state is trivial, since spins 1 and 3 are in the same state and this
state is invariant under the $XY$ interaction. We therefore only
have to consider the $\beta|1\rangle$ part.

The chosen initial state of the spin chain is not unique. We could,
e.g., choose to start with the total system in
$\alpha|011\rangle+\beta|111\rangle$. In this case, the
$|111\rangle$ is invariant and only the transfer of the
$\alpha|0\rangle$ part needs to be considered. At the end of this
section, we discuss additional possibilities.

The iterative transfer scheme of Burgarth et al. consists of a continuous
evolution under the spin-chain Hamiltonian, interrupted by successive applications
of the end-gate operation.
We write the transfer operator as
\begin{equation}\label{iter}
  T_{k}= \prod_{n=1}^{k}W^{23}(c_{n}, d_{n})U^{12}(\tau)
  \label{e.Tk}
\end{equation}
where
\begin{equation}\label{U12t}
 U^{12}(\tau) = e^{-i\tau H_{12}}\otimes I^{3} =
\left (\begin{array}{cccc}
                   1 & 0 & 0 & 0 \\
                   0 & C_{12} & -i S_{12} & 0 \\
                   0 & -i S_{12} & C_{12} & 0 \\
                   0 & 0 & 0 & 1
                 \end{array}
    \right )
\otimes \left ( \begin{array}{cc}
    1 & 0\\
    0 & 1
   \end{array}
    \right )
\end{equation}
represents the evolution of the spin chain and
\begin{equation}\label{psgate}
    W^{23}(c_{n},d_{n})=
    \left ( \begin{array}{cc}
    1 & 0\\
    0 & 1
   \end{array}
    \right )
\otimes \left (\begin{array}{cccc}
                   1 & 0 & 0 & 0 \\
                   0 & d_{n}^{*} & c_{n}^{*} & 0 \\
                   0 & -c_{n} & d_{n} & 0 \\
                   0 & 0 & 0 & 1
                 \end{array}
    \right )
\end{equation}
the end gate operation.
Here, $C_{12} = \cos(\pi J_{12} \tau )$ and
$S_{12} = \sin(\pi J_{12} \tau )$ and $n$ represents the iteration
step.
The parameters $c_n, d_n$ are related by the unitarity
condition $|c_{n}|^{2}+|d_{n}|^{2}=1$. For each step of the
iteration, they are equal to the coefficients of the relevant states
$|010\rangle$ and $|001\rangle$ just before the gate is applied.
Under this condition,
$$
W^{23}(c_{n},d_{n})(c_{n}|010\rangle+d_{n}|001\rangle)=|001\rangle ,
$$
i.e. the transfer to the final state $|001\rangle$ is maximized.

During the $n^{th}$ step, the two coefficients are
\begin{equation}\label{cn}
 c_{n}=-i\frac{S_{12}C^{n-1}_{12}}{\sqrt{1-C_{12}^{2n}}},
 \label{Uini}
\end{equation}
\begin{equation}
\label{dn}
  d_{n}=\sqrt{\frac{1-C_{12}^{2(n-1)}} {1-C_{12}^{2n}}}.
\end{equation}

\subsection{Quantification of transfer}

After $k$ iterations, $|100\rangle$ is transferred to
\begin{equation}
\label{phik}
    |\Psi_{k}\rangle=T_{k}|100\rangle=C_{12}^{k}|100\rangle+
    \sqrt{1-C_{12}^{2k}}|001\rangle.
\end{equation}
Apparently, the transfer increases monotonically with the number of iterations
and can asymptotically approach unity provided $|C_{12}| < 1$.
Writing $F_{k}=\langle001|\Psi_{k}\rangle$
for the overlap of the system with the target state, we find
\begin{equation}\label{Fidelty}
    F_{k}=\sqrt{1-C_{12}^{2k}}.
\end{equation}

Eq. (\ref{e.Tk}) implies that only the spin chain or the end gate are active
at a given time.
If the spin chain interactions are static (not switchable), this can only be
realized approximately if the coupling between the two end-gate qubits
is much stronger than the couplings in the spin chain, $J_{23} \gg J_{12}$.
In the NMR system, we instead refocus the spin-chain interaction during
the application of the end-gate operation to better approximate
the ideal operation
\begin{equation}\label{impsg}
    W^{23}(c_n,d_n)=e^{-i\pi J_{23}t_{n}(\sigma_{x}^{2}\sigma_{x}^{3}+\sigma_{y}^{2}\sigma_{y}^{3})/2}
\end{equation}
where
\begin{equation}\label{endgateimp}
    \tan(\pi J_{23}t_{n})=-ic_n/d_n \, .
\end{equation}

\subsection{Generalization to mixed states}

The IQST algorithm works
also when the spin chain is in a suitable mixed state.
As an example, we choose $\alpha = \beta = \frac{1}{\sqrt{2}}$.
The second
and third qubit can be chosen in any combination of $|0\rangle$ and
$|1\rangle$. Here, we implement all four possibilities in parallel \cite{parallel}
by putting qubits 2 and 3 into the maximally mixed state
$I^{2}\otimes I^{3}$, where $I$ denotes the unit operator
and the upper index labels the qubit.
The sample thus contains an equal number of
molecules with qubits in the states $\alpha |0l\rangle + \beta
|1l\rangle$ with $l =\{ 00, 01, 10, 11 \}$. The traceless part of
the corresponding density operator is \cite{Chuang}
\begin{equation}\label{ini}
    \rho_{ini} = \sum_{l=00}^{11}\sigma_{x}^{1}\otimes(|l\rangle\langle l|).
\end{equation}

If the system is initially in one of the states
$|l\rangle = |01 \rangle, |10 \rangle $, it acquires an overall
phase factor of $-1$ during the transfer.
Combining this with the results of Sec. \ref{IQSTtheory}, we find that
after $k$ iterations, the system is in the state
\begin{equation}\label{rhok}
    \rho_{k} = T_{k} \, \rho_{ini} \, T_{k}^{\dag} = \sqrt{1-F_k^2} \,
    \sigma^{1}_{x} \, I^2I^3+ F_k \, \sigma^{1}_{z}\sigma^{2}_{z}\sigma^{3}_{x}.
\end{equation}

Similarly, when the initial state is chosen as
\begin{equation}\label{inimy}
\rho_{ini} = \sum_{l=00}^{11}\sigma_{y}^{1}\otimes(|l\rangle\langle
l|),
\end{equation}
the algorithm generates the state
\begin{equation}\label{rhok2}
    \rho_{k} = T_{k} \rho_{ini} T_{k}^{\dag} = \sqrt{1-F_k^2} \, \sigma^{1}_{y}I^2I^3+
   F_k \, \sigma^{1}_{z}\sigma^{2}_{z}\sigma^{3}_{y}
\end{equation}
after $k$ iterations.

\section{Implementation}

\begin{figure}
\includegraphics[width=4in]{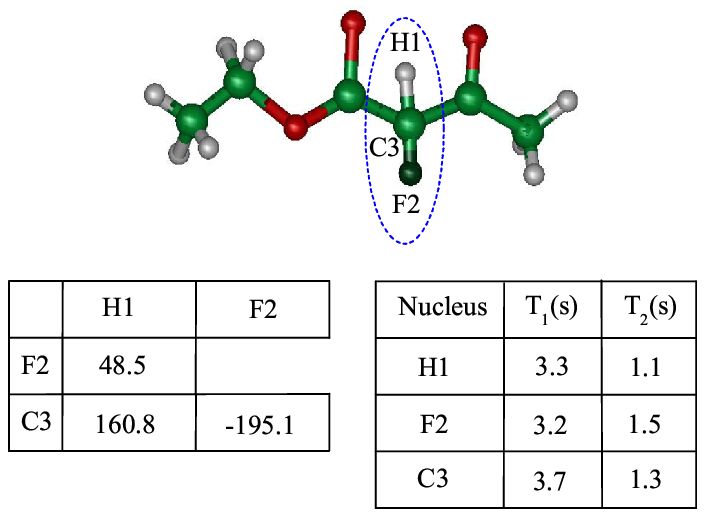}    
\caption{(Color online) The chemical structure of Ethyl
2-fluoroacetoacetate. The three spins in the dashed oval are the
three qubits for implementing IQST. The strengths (in Hz) of the
$J$-couplings between the relevant nuclear spins and the relaxation
times are listed in the left and right tables, respectively.}
\label{2F}
\end{figure}

For the experimental implementation, we chose the $^{1}$H, $^{19}$F,
and $^{13}$C spins of Ethyl 2-fluoroacetoacetate as qubits. The
chemical structure of Ethyl 2-fluoroacetoacetate is shown in Figure
\ref{2F}, where the three qubits are denoted as H1, F2, and C3,
respectively. The strengths of the $J$-couplings are $J_{12}=48.5$
Hz, $J_{23}=-195.1$ Hz and $J_{13}=160.8$ Hz. $T_1$ and $T_2$ values
for these three nuclei are listed in the right table in Figure
\ref{2F}. In the rotating frame, the Hamiltonian of the three- qubit
system is \cite{Chuang,Ernst,CoryPRL99}
\begin{equation}\label{HamCHF}
   H_{NMR}=\frac{\pi}{2}(J_{12}\sigma^1_z\sigma^2_z
   +J_{23}\sigma^2_z\sigma^3_z+J_{13}\sigma^1_z\sigma^3_z).
\end{equation}

\begin{figure}
\includegraphics[width=3in]{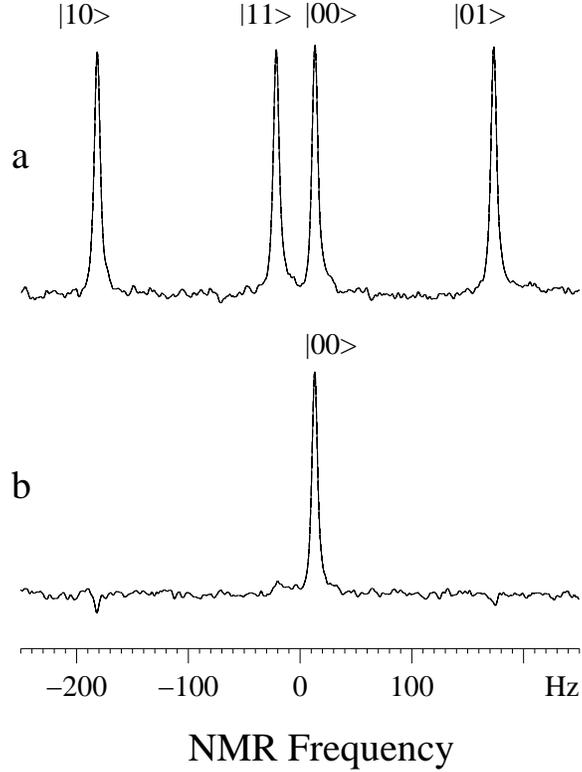}    
\caption{(a) $^{13}$C NMR spectrum obtained by applying a selective
readout pulse to the system in its thermal equilibrium state.
The four resonance lines correspond to specific states of the spin-chain qubits
H1 and F2, as
indicated by the labels above the resonance lines. The assignment
takes into account that $J_{13}>0$ and $J_{23}<0$.
(b) $^{13}$C NMR spectrum of the state $|00\rangle(|0\rangle-|1\rangle)/\sqrt{2}$,
which was obtained by applying a $[\pi/2]_{y}^{3}$ pulse to $|000\rangle$.}
\label{inistate}
\end{figure}

The sample consisted of a 3:1 mixture of unlabeled  Ethyl
2-fluoroacetoacetate and d6-acetone. Molecules with a $^{13}$C
nucleus at position 2, which we used as the quantum register, were
therefore present at a concentration of about $1 \%$. They were
selected against the background of molecules with $^{12}$C nuclei by
measuring the $^{13}$C signal. We chose H1 as the input qubit and C3
as the target qubit. Figure \ref{inistate} (a) shows the $^{13}$C NMR
spectrum obtained by applying a readout pulse to the system in its
thermal equilibrium state. Each of the resonance lines is associated
with a specific spin state of qubits 1 and 2.

\subsection{Initial state preparation}

The initial pseudo-pure state $|000\rangle$  is prepared by spatial
averaging \cite{spatial}. The following radio-frequency (rf) and
magnetic field gradient pulse sequence transforms
the system from the equilibrium state
\begin{equation}\label{equ}
  \rho_{eq}=\gamma_{1}\sigma_{z}^{1}+ \gamma_{2}\sigma_{z}^{2}+\gamma_{3}\sigma_{z}^{3}
\end{equation}
to $|000\rangle$:
$[\phi_{1}]_{y}^{1}-[\phi_{2}]_{y}^{2}-[grad]_{z}-[\pi/2]^{1}_{x}-[1/2J_{13}]
-[-\pi/2]^{1}_{y}-[\pi/4]^{3}_{x}-[-1/2J_{23}]-[-\pi/4]^{3}_{y}-[grad]_{z}
-[\pi/4]^{1}_{x}-[1/2J_{13}]-[-\pi/4]^{1}_{y}-[grad]_{z}$.
Here $\gamma_{1}$, $\gamma_{2}$ and $\gamma_{3}$
denote the gyromagnetic ratios of H1, F2, and C3, respectively, and
$\cos \phi_{1}=2\gamma_{3}/\gamma_{1}$, and $\cos
\phi_{2}=\gamma_{3}/2\gamma_{2}$. $[grad]_{z}$ denotes a gradient
pulse along the $z$- axis. $[\pi/2]_{x}^{1}$ denotes a $\pi/2$ pulse
along the $x$- axis acting on the H1 qubit. Overall phase factors
have been ignored.

The coupled-spin evolution between two spins, for
instance, $[1/2J_{13}]$, can be realized by the pulse sequence
$1/4J_{13}-[\pi]^{2}_{y} -1/4J_{13}-[-\pi]^{2}_{y}$, where
$1/4J_{13}$ denotes the evolution caused by $H_{NMR}$ for a time
$1/4J_{13}$ \cite{ZZcouple}.

The target state can be prepared directly from the state $|000\rangle$
by applying a $[\pi/2]^{3}_{y}$ pulse.
It corresponds to $|00\rangle(|0\rangle-|1\rangle)/\sqrt{2}$,
i.e. to transverse magnetization
of the target spin, with the first two qubits in state $\vert 00
\rangle$. If we measure the free induction decay (FID) of this state
and calculate the Fourier transform of the signal, we obtain the
spectrum shown in Figure \ref{inistate} (b). This spectrum serves as
the reference to which we scale the data from the IQST experiment.

  The input state for the IQST is
$|\Psi_{in}\rangle=|\psi(\theta)\rangle|00\rangle$.
We generate this state by rotating H1 by an angle
$\theta$ around the $y$-axis:
$|\Psi_{in}\rangle = e^{i\theta\sigma^{1}_{y}/2} |000\rangle$.
After $k$ iterations of the IQST algorithm, $|\Psi_{in}\rangle$ is transferred to
\begin{equation}
\label{phikout}
    T_{k}|\Psi_{in}\rangle=
    [(1-F_k)\cos(\theta/2)|0\rangle-\sqrt{1-F_{k}^{2}}\sin(\theta/2)|1\rangle]|00\rangle+|00\rangle
    F_{k}|\psi(\theta)\rangle .
\end{equation}
Here, we have used Eqs. (\ref{phik}-\ref{Fidelty}) and assumed $C_{12}\geq 0$,
without loss of generality. Hence the state transfer can be observed
through measuring carbon spectra.

For the mixed input state, $\rho_{ini}$ [Eq. (\ref{inimy})] can be
generated from $\rho_{eq}$ through the pulse sequence \cite{Tseng}
\begin{eqnarray}\label{inip}
[\frac{\pi}{2}]_{x}^{3}-[\frac{\pi}{2}]_{x}^{2}-[grad]_{z}-[\frac{\pi}{2}]_{x}^{1}.
\end{eqnarray}
\subsection{Effective XY-interactions}

The IQST algorithm requires XY interactions, while the natural
Hamiltonian contains ZZ couplings. To convert the ZZ interactions
into XY type, we decompose the evolution
$e^{-i\varphi(\sigma_{x}^{k}\sigma_{x}^{l}+\sigma_{y}^{k}\sigma_{y}^{l})}$
into
$e^{-i\varphi\sigma_{x}^{k}\sigma_{x}^{l}}e^{-i\varphi\sigma_{y}^{k}\sigma_{y}^{l}}$
\cite{cory07} using
$[\sigma_{x}^{k}\sigma_{x}^{l},\sigma_{y}^{k}\sigma_{y}^{l}]=0$,
where $\varphi$ denotes an arbitrary real number. These
tranformations can be implemented by a combination of
radio-frequency pulses and free evolutions under the $J$-couplings:
\cite{DuPRA03}.
\begin{equation}\label{XX}
e^{-i\varphi\sigma_{x}^{k}\sigma_{x}^{l}}= e^{\pm
i\pi\sigma_{y}^{k}/4}e^{\pm i\pi\sigma_{y}^{l}/4}
e^{-i\varphi\sigma_{z}^{k}\sigma_{z}^{l}} e^{\mp
i\pi\sigma_{y}^{k}/4}e^{\mp i\pi\sigma_{y}^{l}/4}
\end{equation}
\begin{equation}\label{YY}
e^{-i\varphi\sigma_{y}^{k}\sigma_{y}^{l}}= e^{\pm
i\pi\sigma_{x}^{k}/4}e^{\pm i\pi\sigma_{x}^{l}/4}
e^{-i\varphi\sigma_{z}^{k}\sigma_{z}^{l}} e^{\mp
i\pi\sigma_{x}^{k}/4}e^{\mp i\pi\sigma_{x}^{l}/4} \, .
\end{equation} 

\begin{figure}
\includegraphics[width=6in]{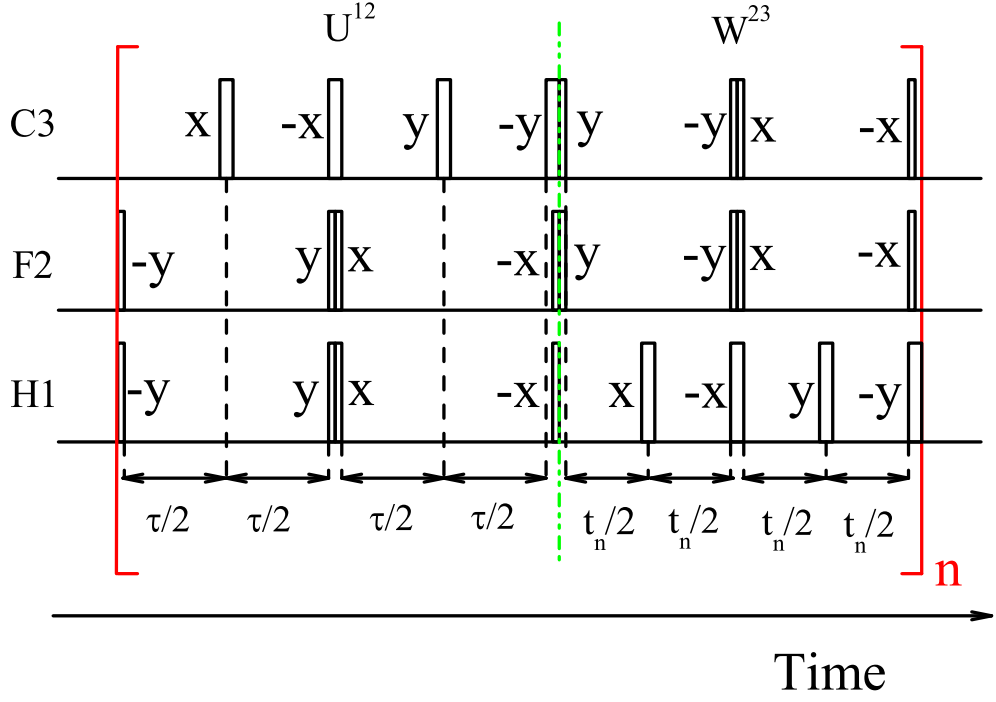}    
\caption{(Color online) Pulse sequence for implementing the IQST.
The two blocks that implement $U^{12}(\tau)$ and $W^{23}(c_n,d_n)$
are separated by the dash-dotted line and "$[...]_{n}$" indicates
iteration $n$. The delays $t_n$ are given by Eq. (\ref{endgateimp}).
The narrow rectangles denote $\pi/2$ pulses, and the wide ones
denote $\pi$ pulses, where $x$, $-x$, $y$, or $-y$ denote the
direction along which the pulse is applied. The $\pi$ pulses are
applied in pairs with opposite phases to reduce experimental errors
\cite{Fang}. The durations of the pulses are so short that they can
be ignored. } \label{pulseq}
\end{figure}

Figure \ref{pulseq} shows the complete pulse sequence for
implementing the IQST, starting from $|\Psi_{in}\rangle$. The
subscript $n$ indicates that the pulses in the square brackets have
to be repeated for every iteration. The duration of each $W^{23}$ segment
varies, since $t_{n}=-\arctan(ic_{n}/d_{n})/\pi J_{23}$.

  For the initial state $\rho_{ini}$ in Eq. (\ref{ini}), the propagators
$n$ can be simplified: since the density operator commutes with
$\sigma^{1}_{x}\sigma^{2}_{x}$ and $\sigma^{2}_{y}\sigma^{3}_{y}$ at
all times, it is sufficient to generate the propagator
$$
e^{-i\pi J_{23}t_n\sigma^{2}_{x}\sigma^{3}_{x}/2}e^{-i\pi
J_{12}\tau\sigma^{1}_{y}\sigma^{2}_{y}/2} .
$$
Similarly, for the initial state in Eq. (\ref{inimy}), iteration $n$
can be replaced by $e^{-i\pi
J_{23}t_n\sigma^{2}_{y}\sigma^{3}_{y}/2}e^{-i\pi
J_{12}\tau\sigma^{1}_{x}\sigma^{2}_{x}/2}$. We use these simplified
versions to shorten the duration of the experiment and thereby
increase the fidelity.

\subsection{Results for state transfer}
\begin{figure}
\includegraphics[width=3in]{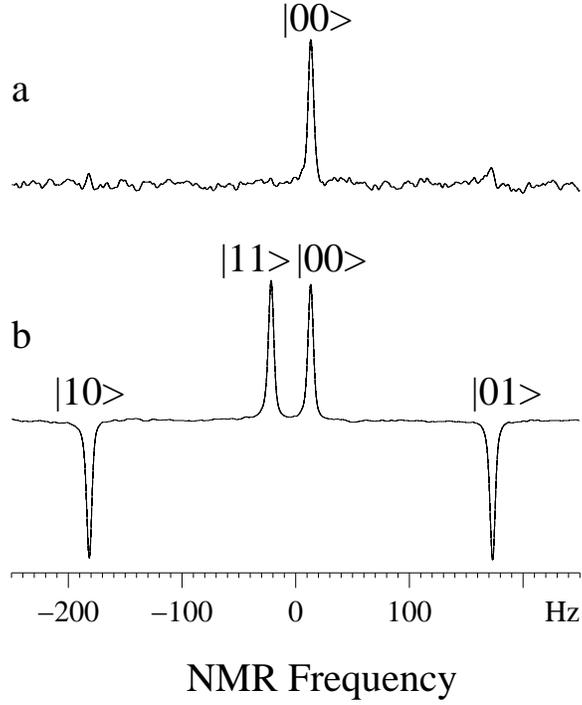}
\caption{Experimental results for quantum state transfer with $\tau=1/2J_{12}$.
The initial states are $[|0\rangle -|1\rangle]|00\rangle/\sqrt{2}$ and  $\sigma_{y}^{1}$, corresponding to figures (a) and (b), respectively.
In the first experiment, the receiver phase was set to $x$, in the second experiment
it was set to $y$.}
\label{onepure}
\end{figure}

  When $\tau=1/2J_{12}$, the transfer can be implemented in a single step
with a theoretical fidelity of $100\%$.
The state transfer
from H1 to C3 can be observed by measuring $^{13}$C spectra.
The
experimental result for
$|\Psi_{in}\rangle=|\psi(\pi/4)\rangle|00\rangle$ is shown in Figure
\ref{onepure} (a).
Comparing with Figure \ref{inistate} (b)
one finds that the output state is
$|00\rangle(|0\rangle-|1\rangle)/\sqrt{2}$, i.e., the state
$|\psi(\pi/4)\rangle$ is transferred from H1 to C3.

Figure \ref{onepure} (b), show the corresponding result for the
transfer of $\sigma_{y}^{1}$ from H1 to C3 in a single step,
with qubits 2 and 3 initially in the completely mixed state.
For this experiment, the receiver phase was shifted by $\pi/2$
with respect to the upper spectrum.
Since this experiment implements the transfer for all possible
states of the other qubits in parallel, we observe four resonance lines
corresponding to the states $\{00, 01, 10, 11\}$ of qubits 1 and 2.
For the states with odd parity, the transfer adds an overall phase factor
of -1, which is directly visible as a negative amplitude in the spectrum.

\begin{figure}
\includegraphics[width=6in]{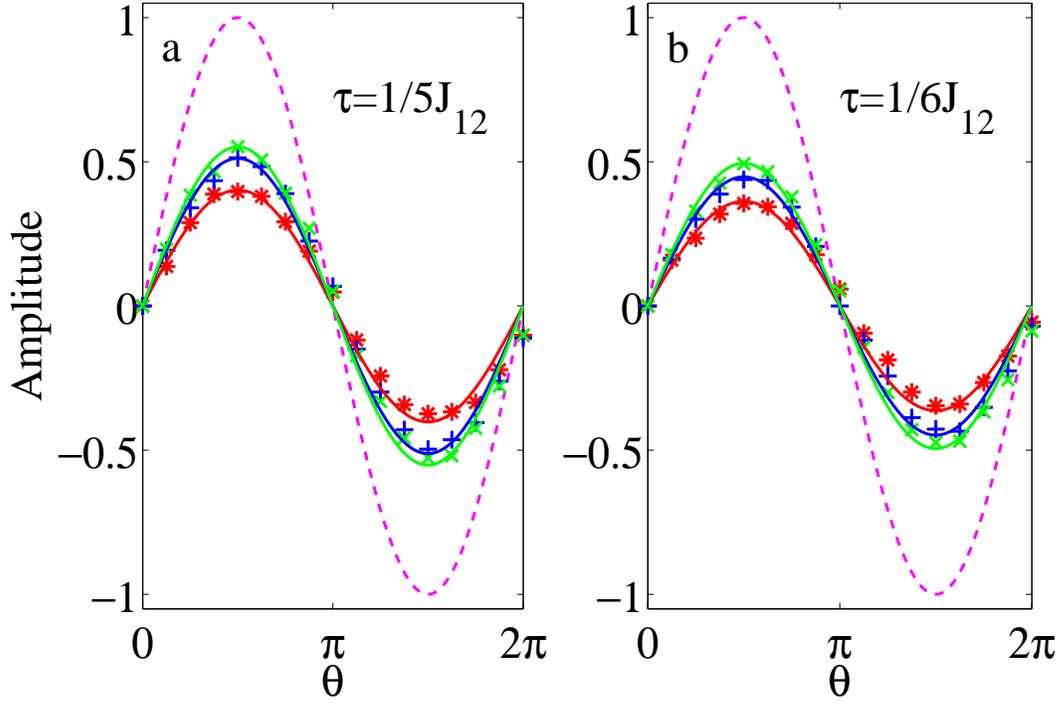}
\caption{(Color online) Experimental results for demonstrating the
IQST when the initial state is $[\cos(\theta/2)|0\rangle
-\sin(\theta/2)|1\rangle]|00\rangle$. Two cases for $\tau=1/5J_{12}$
and $\tau=1/6J_{12}$ are shown in Figures (a) and (b). For each case
three iterations are implemented. The experimental data after the
completion of iteration 1, 2, and 3 are marked by "*", "+", and
"$\times$", respectively. The data can be fitted as sin functions of
which amplitudes represent the measured fidelities experimentally.
The dashed curves show $\sin(\theta)$.} \label{tau56}
\end{figure}

   To demonstrate that iterative transfer works for a range of coupling
strengths or (equivalently) evolution periods, we chose
$\tau=1/5J_{12}$ and $\tau=1/6J_{12}$. For the case of pseudo-pure
input states, three iterations are implemented for either case. When
$\theta$ changes from $0$ to $2\pi$ the experimental results
obtained from these transfer experiments are summarized in Figure
\ref{tau56}, where the vertical axis denotes the amplitude of the
NMR spectrum.
For each input state the amplitude
increases with the number of iterations.
The increase of the
amplitude shows the increase of the fidelity for the state transfer.
The dependence on the input state parameter $\theta$ has the
expected $\sin (\theta)$ dependence.

\begin{figure}
\includegraphics[width=4in]{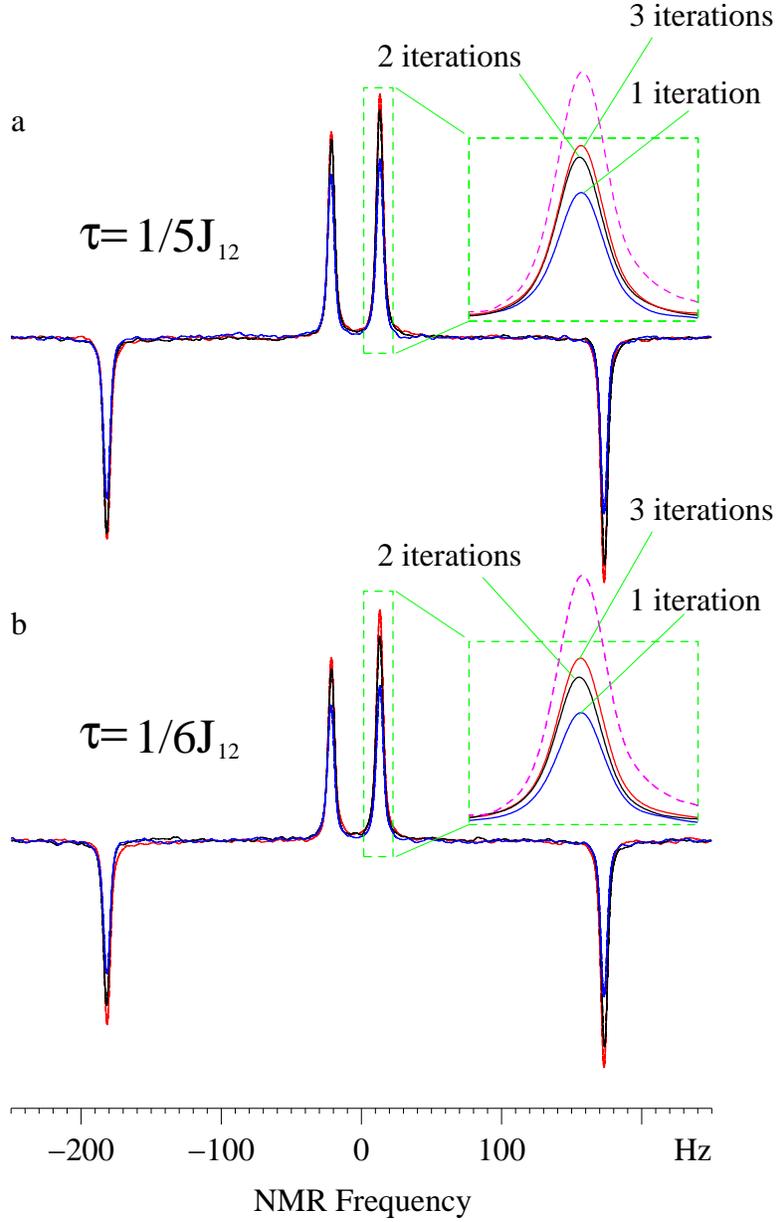}
\caption{(Color online)  $^{13}$C NMR spectra demonstrating the IQST
of the state $\sigma_{y}^{1}$ for $\tau=1/5J_{12}$ and
$\tau=1/6J_{12}$. For each case, the spectra after the completion of
iteration 1, 2, and 3 are shown as the blue, black and red curves,
respectively. The resonance lines corresponding to the $\vert 00
\rangle$ state of the spin chain are enlarged in the inset. The
dashed curves are the corresponding sections of the reference
spectrum in Figure \ref{inistate} (a).
} \label{Res}
\end{figure}

The experimental data obtained for the mixed input states are summarized in
Figures \ref{Res} (a) and (b), for $\tau=1/5J_{12}$ and
$\tau=1/6J_{12}$, respectively. The positive lines indicate that the
transfer occurs with positive sign if qubits 1 and 2 are in state
$|00\rangle$ or $|11\rangle$, and with negative sign for the states
$|01\rangle$ or $|10\rangle$, in agreement with Eq. (\ref{rhok2}).
Obviously the amplitude of the signals increases with the number of
iterations.
According to Eq. (\ref{rhok2}) the increase of the amplitudes
is a direct measure for the progress of the quantum state transfer.

\section{Discussion and Conclusion}

Our results clearly demonstrate the validity of the iterative state
transfer algorithm of Burgarth et al. In principle, it is possible
to iterate the procedure indefinitely, always improving the fidelity
of the transfer. In practice, every iteration also increases the
amount of signal loss, either through decoherence or through
experimental imperfections.

\begin{figure}
\includegraphics[width=5in]{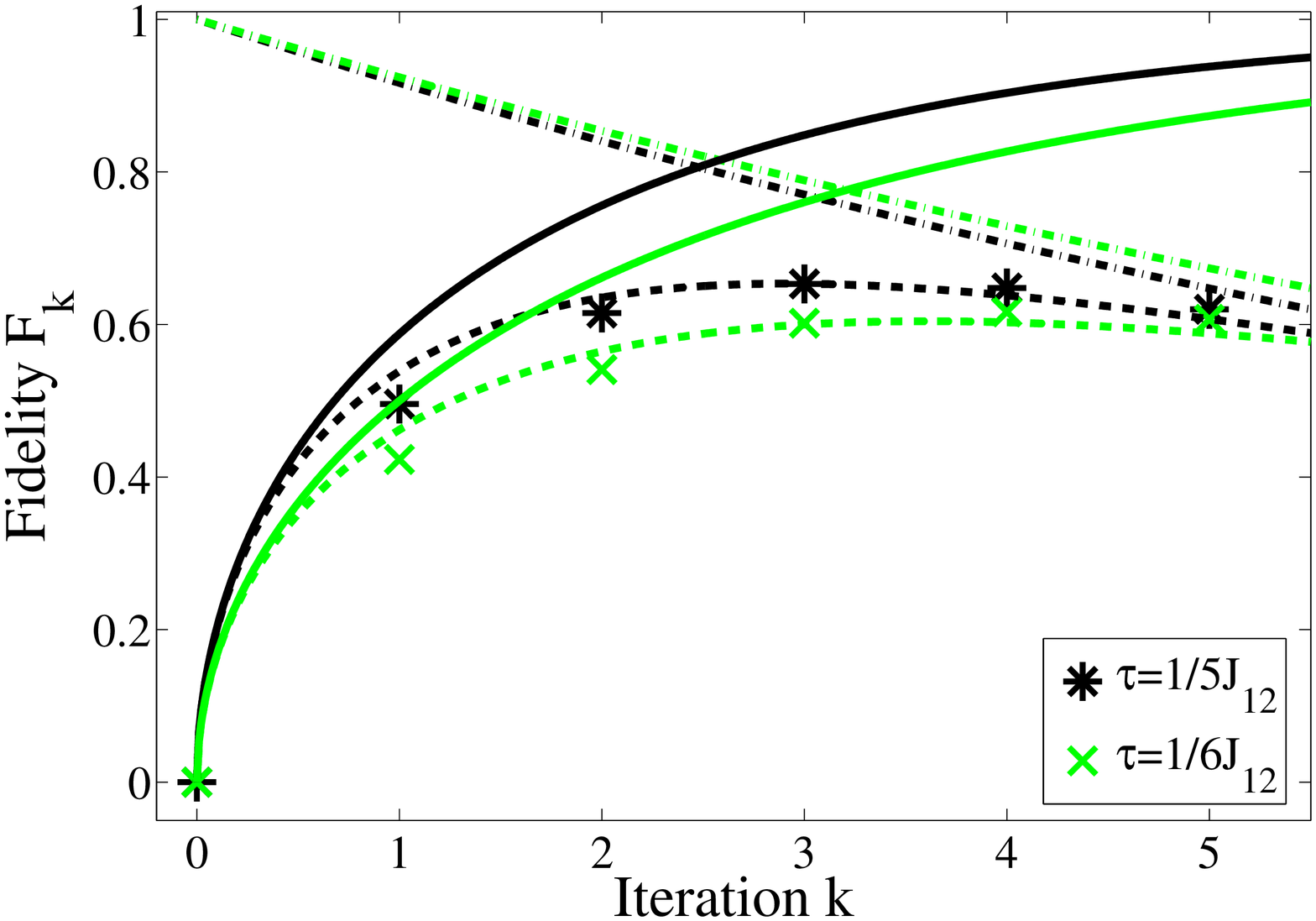}     
\caption{(Color online) Experimentally measured fidelity of the
iterative state transfer as a function of the number of iteration
steps when $\tau=1/5J_{12}$ and $\tau=1/6J_{12}$. The experimental
data are fitted to the function $F_{k}e^{-kr}$ with $r=0.087$ and
$0.079$ for the two cases, respectively. The two solid curves
represent the theoretical fidelities $F_{k}$ for ideal conditions,
and the two dash-dotted curves correspond to
 $e^{-kr}$.
The dark and light curves correspond to the cases of $\tau=1/5J_{12}$
and $\tau=1/6J_{12}$, respectively.} \label{tau56x}
\end{figure}

According to Eq. (\ref{rhok2}), the fidelity of the transfer is
\begin{equation}\label{expzzx}
 F_{k}=|Tr[(\sigma^{1}_{z}\sigma^{2}_{z}\sigma^{3}_{y})\rho_{k}]| .
\end{equation}
The experimental measurement corresponds to a summation of the
amplitudes of the resonance lines. We normalized the experimental
values to the amplitudes obtained by direct preparation of the
target states [see Figure \ref{inistate} (a)]. In Figure
\ref{tau56x}, we show the experimentally measured fidelities of the
transfer of the state $\sigma_{y}$ for 1-5 iterations. As expected,
the experimental data points are below the theoretical curves (full
lines).

The experimental points can be fitted quite well if we include a decay parameter
for each iteration.
The dashed curves in Figure \ref{tau56x} represent the function
$F_{k}e^{-kr}$ with $r=0.087$ and $r=0.079$ for
$\tau=1/5J_{12}$ and $\tau=1/6J_{12}$, respectively.
Each iteration thus adds imperfections (experimental plus decoherence)
of about 8 \%.
Larger numbers of iterations are meaningful only if this error rate
can be reduced.

In conclusion, we have implemented the iterative quantum state transfer in a three
qubit NMR quantum information processor.
The result shows that it is indeed possible to accumulate the quantum state
at the end of a Heisenberg spin chain, whose couplings are always active.

\section{Acknowledgment}

We thank Prof. Jiangfeng Du for helpful discussions. This work is
supported by the Alexander von Humboldt Foundation, the DFG through
Su 192/19-1, and the Graduiertenkolleg No. 726.



\begin{thebibliography}{}
\bibitem{books} M. A. Nielsen and
                I. L. Chuang, {\it Quantum Computation and Quantum Information} (Cambridge
                University Press, Cambridge, England, 2000);
                {\it The Physics of Quantum Information}, edited by
                D. Bouwmeester, A. Ekert, and A. Zeilinger (Springer, Berlin,
                2000).

\bibitem{Bose03} S. Bose, Phys. Rev. Lett. {\bf 91}, 207901 (2003).

\bibitem{PST} M. Christandl, N. Datta, A. Ekert, and A. J. Landahl, Phys. Rev. Lett. {\bf 92},
              187902(2004); M. Christandl, N. Datta, T. C. Dorlas, A. Ekert,
              A. Kay, and A. J. Landahl, Phys. Rev. A {\bf 71},
                032312(2005).


\bibitem{thesis}D. Burgarth, arXiv: 0704.1309 [quant-ph];
                     arXiv: 0706.0387 [quant-ph];
            D. L. Feder, Phys. Rev. Lett. {\bf 97}, 180502 (2006);
            A. Kay, {\it ibid}. {\bf 98}, 010501 (2007);
                Phys. Rev. A {\bf 73}, 032306 (2006);
                X.-F. Qian, Y. Li, Y. Li, Z. Song, and C. P. Sun,
                {\it ibid}. {\bf 72}, 062329 (2005);
                P. Karbach and J. Stolze, {\it ibid}. {\bf 72}, 030301(R)
                (2005);
                M.-H. Yung, {\it ibid}. {\bf 74}, 030303(R) (2006);
                P. K. Gagnebin, S. R. Skinner, E. C. Behrman, and J. E.
                Steck, {\it ibid}. {\bf 75}, 022310 (2007);
                V. Kostak, G. M. Nikolopoulos, and I. Jex, {\it ibid}. {\bf 75}, 042319 (2007);
                O. Romero-Isart, K. Eckert, and A. Sanpera, {\it ibid}. {\bf 75}, 050303(R)
                (2007);
                A. Bayat and V. Karimipour, {\it ibid}. {\bf 75}, 022321
                (2007);
                A. W\'{o}jcik, et al., {\it ibid}. {\bf 75}, 022330
                (2007);
                 K. Eckert, O. Romero-Isart, and A. Sanpera, New J. Phys. {\bf 9}, 155
                 (2007);
                 P. Cappellaro, C. Ramanathan, D. G. Cory,
                 arXiv:0706.0342  [quant-ph].

\bibitem{Bose} D. Burgarth, V. Giovannetti, S. Bose,
                Phys. Rev. A {\bf 75}, 062327 (2007).

\bibitem{swap} Z. L. Madi, R. Br$\ddot{u}$schweiler, and R. R. Ernst, J. Chem. Phys. {\bf 109},
                10603 (1998).

\bibitem{telep} C. H. Bennett, G. Brassard, C. Cr$\acute{e}$peau, R.
                Jozsa, A. Peres, and W. K. Wootters, Phys. Rev. Lett. {\bf 70},
                1895 (1993);
                D. Boschi, S. Branca, F. D. Martini, L. Hardy, and S.
                 Popescu, {\it ibid}. {\bf 80}, 1121 (1998);
                D. Bouwmeester, J. Pan, K. Mattle, M. Eibl, H. Weinfurter,
                and A. Zeilinger, Nature (London) {\bf 390}, 575
                (1997);
                M. A. Nielsen, E. Knill, and R. Laflamme, {\it ibid}. {\bf 396}, 52
                    (1998).

\bibitem{parallel} E. Knill and R. Laflamme, Phys. Rev. Lett. {\bf 81}, 5672 (1998);
                A. Datta, S. T. Flammia, and C. M. Caves, Phys.
                Rev. A {\bf 72}, 042316(2005);
                R. Stadelhofer, D. Suter, and W. Banzhaf,
                        {\it ibid}. {\bf71}, 032345 (2005);
                G. L. Long, and L. Xiao, J. Chem. Phys. {\bf 119},
                8473 (2003).

\bibitem{Chuang} I. L. Chuang, N. Gershenfeld, M. G. Kubinec, and D. W. Leung,
                    Proc. R. Soc. London, Ser. A {\bf 454}, 447 (1998).

\bibitem{Ernst} R. R. Ernst, G.Bodenhausen, and A.Wokaum, {\it Principles of
                Nuclear Magnetic Resonance in One and Two Dimensions} (Oxford
                University Press, Oxford, 1987).

\bibitem{CoryPRL99} S. Somaroo, C. H. Tseng, T. F. Havel, R. Laflamme, and D. G.
                        Cory, Phys. Rev. Lett. {\bf 82}, 5381
                        (1999).

\bibitem{spatial} D. G. Cory, M. D. Price, and T. F. Havel, Physica D {\bf 120}, 82
                  (1998);
                  J.-F. Zhang, G. L. Long, Z.-W. Deng, W.-Z. Liu, and Z.-H.
                  Lu, Phys. Rev. A {\bf 70}, 062322 (2004);
                  X.-H. Peng, X.-W. Zhu, X.-M. Fang, M. Feng, X.-D.
                  Yang, M.-L. Liu, and K.-L. Gao,
                  arXiv:quant-ph/0202010.

\bibitem{ZZcouple}  L. M. K. Vandersypen, M. Steffen, M. H. Sherwood, C. S. Yannoni, G. Breyta, and I. L.
                  Chuang, Appl. Phys. Lett. {\bf 76}, 646 (2000);
                  L. M. K. Vandersypen and I. L. Chuang Rev. Mod. Phys. {\bf 76},
                  1037 (2004);
                  N. Linden, $\bar{E}$. Kup$\check{c}$e, and R. Freeman, Chem. Phys. Lett. {\bf 311},
                    321 (1999);
                  X.-H. Peng, X.-W. Zhu, M. Fang, M.-L. Liu, and K.-L. Gao,
                  Phys. Rev. A {\bf 65}, 042315 (2002);
                    R. Somma, G. Ortiz, J. E. Gubernatis, E. Knill, and R.
                    Laflamme, {\it ibid}. {\bf 65}, 042323 (2002).

\bibitem{Tseng} C. H. Tseng, S. Somaroo, Y. Sharf, E. Knill, R. Laflamme, T. F.
                 Havel, and D. G. Cory, Phys. Rev. A {\bf61},
                 012302(1999).

\bibitem{cory07} J. S. Hodges, P. Cappellaro, T. F. Havel, R. Martinez, and D. G.
                Cory,  Phys. Rev. A {\bf 75}, 042320 (2007).

\bibitem{DuPRA03} S. S. Somaroo, D. G. Cory and T. F. Havel, Phys. Lett.
              A {\bf 240}, 1 (1998);
            M. D. Price, S. S. Somaroo, A. E. Dunlop, T. F. Havel, and D. G.
               Cory, Phys. Rev. A {\bf 60}, 2777 (1999);
            J.-F. Du, H. Li, X.-D. Xu, M.-J. Shi, J.-H Wu,
              X.-Y Zhou, and R.-D. Han, {\it ibid}. {\bf 67}, 042316
              (2003);
            J.-F. Zhang, G. L. Long, W. Zhang, Z.-W. Deng, W.-Z. Liu, and
                  Z.-H. Lu, {\it ibid}. {\bf 72}, 012331 (2005);
            J.-F. Zhang, X.-H. Peng, D. Suter, {\it ibid}. {\bf 73}, 062325
                    (2006).

\bibitem{Fang}  X.-M. Fang, X.-W. Zhu, M. Feng, X.-A. Mao, and F. Du, Phys.
                Rev. A {\bf 61}, 022307 (2000).







\end{thebibliography}
\end{document}